\documentclass[sigconf]{acmart}
\AtBeginDocument{%
  }

\copyrightyear{2025}
\acmYear{2025}
\setcopyright{acmlicensed}\acmConference[MM '25]{Proceedings of the 33rd ACM International Conference on Multimedia}{October 27--31, 2025}{Dublin, Ireland}
\acmBooktitle{Proceedings of the 33rd ACM International Conference on Multimedia (MM '25), October 27--31, 2025, Dublin, Ireland}
\acmDOI{10.1145/3746027.3755828}
\acmISBN{979-8-4007-2035-2/2025/10}

\begin{document}

\title{Robust Photo-Realistic Hand Gesture Generation: \\
from Single View to Multiple View}

\author{Qifan Fu}
\authornote{Corresponding author.}
\email{q.fu@qmul.ac.uk}
\orcid{0000-0003-3505-9865}
\affiliation{%
  \institution{Digital Environment Research Institute \& School of Electronic Engineering and Computer Science, Queen Mary University of London}
  \city{London}
  \country{UK}
}

\author{Xu Chen}
\affiliation{%
  \institution{Digital Environment Research Institute, Queen Mary University of London}
  \city{London}
  \country{UK}}
\affiliation{%
  \institution{Department of Medicine, University of Cambridge}
  \city{Cambridge}
  \country{UK}}

\author{Muhammad Asad}
\affiliation{%
  \institution{Digital Environment Research Institute, Queen Mary University London}
  \city{London}
  \country{UK}
}

\author{Shanxin Yuan}
\affiliation{%
  \institution{School of Electronic Engineering and Computer Science \& Digital Environment Research Institute, Queen Mary University of London}
  \city{London}
  \country{UK}
}

\author{Changjae Oh}
\affiliation{%
  \institution{School of Electronic Engineering and Computer Science, Queen Mary University of London}
  \city{London}
  \country{UK}}

\author{Gregory Slabaugh}
\affiliation{%
  \institution{Digital Environment Research Institute, Queen Mary University London}
  \city{London}
  \country{UK}}

\renewcommand{\shorttitle}{Robust Photo-Realistic Hand Gesture Generation: 
from Single View to Multiple View}

\begin{abstract}
High-fidelity hand gesture generation represents a significant challenge in human-centric generation tasks. Existing methods typically employ a single-view mesh-rendered image prior to enhancing gesture generation quality. However, the spatial complexity of hand gestures and the inherent limitations of single-view rendering make it difficult to capture complete gesture information, particularly when fingers are occluded. The fundamental contradiction lies in the loss of 3D topological relationships through 2D projection and the incomplete spatial coverage inherent to single-view representations. Diverging from single-view prior approaches, we propose a multi-view prior framework, named Multi-Modal UNet-based Feature Encoder (MUFEN), to guide diffusion models in learning comprehensive 3D hand information. Specifically, we extend conventional front-view rendering to include rear, left, right, top, and bottom perspectives, selecting the most information-rich view combination as training priors to address occlusion. This multi-view prior with a dedicated dual stream encoder significantly improves the model's understanding of complete hand features. Furthermore, we design a bounding box feature fusion module, which can fuse the gesture localization features and multi-modal features to enhance the location-awareness of the MUFEN features to the gesture-related features. Experiments demonstrate that our method achieves state-of-the-art performance in both quantitative metrics and qualitative evaluations. The source code is available at \url{https://github.com/fuqifan/MUFEN}.
\end{abstract}



\begin{teaserfigure}
  \centering
  \includegraphics[width=1\textwidth]{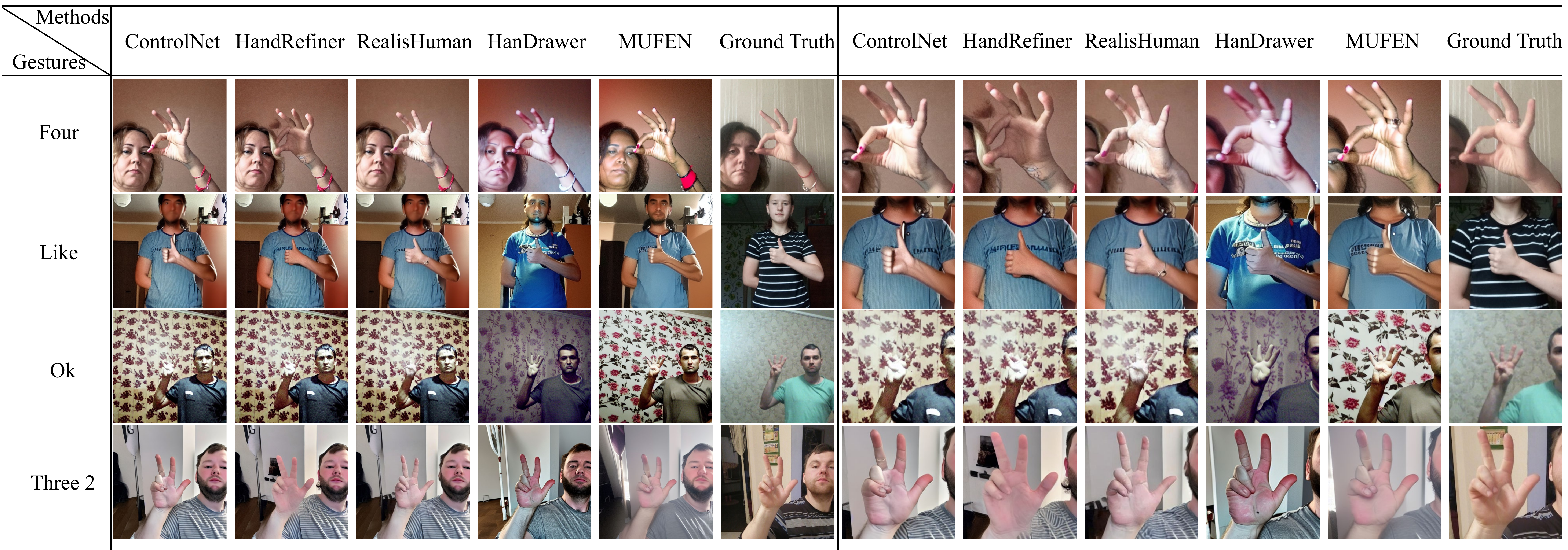}
  \caption{Photo-realistic hand gesture generation via various methods. Our method, named MUFEN, improves the realism and accuracy of gesture generation by fusing multi-view and multi-modal features of the gesture.}
  \label{fig:1}
\end{teaserfigure}


\maketitle

\section{Introduction}
Human-centric image and video generation has been a long-standing focus in the field of computer vision~\cite{zhu2024mole,ma2024follow}. Despite remarkable advancements in diffusion models that have enabled more realistic human generation~\cite{wang2025realishuman} and improved pose control accuracy~\cite{zhang2023adding,li2024controlnet++}, achieving natural and precise hand gesture generation remains a critical challenge, with persistent issues of unnatural occurrences such as finger multiplicity or deficiency, as shown in Figure~\ref{fig:1}. This challenge primarily arises from two inherent characteristics: Firstly, the hand is highly flexible, capable of diverse gestures - although occupying minimal spatial proportion relative to the full body, the hands possess extraordinary degrees of freedom and movement complexity. Secondly, severe occlusion phenomena, particularly inter-finger occlusions, create substantial obstacles for models to learn comprehensive hand representations, potentially leading to incomplete hand modelling through partial observation.

The challenge of diverse hand gestures can be addressed through the enhancement of dataset content, model structure, and training algorithm. Upper-body-focused datasets are carefully designed to emphasize hand gesture representations~\cite{kapitanov2024hagrid, lin2023one, duarte2021how2sign}. Notably, HaGRID~\cite{kapitanov2024hagrid} – originally designed for gesture recognition – has emerged as the most popular dataset for hand generation tasks. Beyond specialized gesture dataset construction, numerous methods have developed dedicated architectures or algorithms to enhance gesture generation quality. These approaches are broadly categorized into single-stage generation, and multi-stage inpainting methodologies.
For single-stage generation, researchers primarily design dedicated encoders to extract rich features from 3D meshes or depth image priors, thereby guiding generation models to synthesize high-fidelity hand gestures~\cite{narasimhaswamy2024handiffuser}. Multi-stage approaches typically employ specialized generative models to refine hand regions after initial full-image generation by general-purpose models, implementing targeted corrections through cascaded networks~\cite{lu2024handrefiner}.

Although existing methods have made progress in improving the quality of gesture generation, they typically rely on mesh renderings from a single viewpoint as prior~\cite{lu2024handrefiner, wang2025realishuman}. This limitation prevents the model from acquiring sufficient 3D information of the hand gestures, especially in cases of self-occlusion, where critical structural information is often missing.
As illustrated in Figure~\ref{fig:2} with the thumb obscured, complex gestures often involve severe occlusions, particularly in the fingers, where occluded parts result in ambiguities, making it difficult for the model to infer accurate finger configurations. Such inaccurate modelling severely hinders the learning of complete hand representations, leading to significant distortions during gesture generation, such as missing or multiple fingers.

To address occlusion issues in gestures and enable the model to learn comprehensive hand modelling information, we rendered the meshes from front, rear, left, right, top, and bottom perspectives to supplement the occlusion information in a single view. We then designed a Multi-Modal UNet-based Feature Encoder (MUFEN) to extract multi-view multi-modal features of hand gesture. Specifically, MUFEN uses dual stream rendering encoder for the rendering meshes from different perspectives to extract multi-view features. Deep features and linguistic labeling features are extracted using a trainable deep encoder and a frozen CLIP, respectively. Additionally,  to enhance the spatial information of multimodal hand features extracted by MUFEN, a bounding box feature fusion module is proposed for enhancing multimodal gesture features with gesture localization features. This allows the MUFEN features to be fused directly with the features of the diffusion model, without the need for separate adjustments to the fusion region like zero padding strategies~\cite{fu2025handrawer} that have appeared in the literature.

In summary, the contributions of this paper are as follows:
\begin{itemize}
    \item We utilize multi-view (including front, rear, left, right, top, and bottom perspectives) gesture rendering mesh priors to supplement occluded hand information in single-view scenarios.

    \item We design a multi-modal UNet-based feature encoder named MUFEN with dual stream rendering encoder to extract and fuse the multi-view multi-modal features of the gesture region, to enhance the model's ability to learn comprehensive hand modelling information.
    
    \item We leverage gesture localization features from a bounding box to enhance MUFEN features by a trainable bounding box feature fusion module, enabling MUFEN features to be fused directly with the diffusion model features without the need for fusion region adjustment.
    
    \item Simulations demonstrate that the proposed MUFEN achieves state-of-the-art (SOTA) performance in both quantitative and qualitative analyses.

\end{itemize}

\section{Related Work}
\subsection{Controllable Human-Centric Generation}
Controllable human-centric image or video generation is a popular task. Recently, ControlNet~\cite{zhang2023adding} and T2I-Adapter~\cite{mou2024t2i} enable conditional human pose generation by incorporating control signals such as skeletons, sketches, and depth maps.
And then, ControlNet++~\cite{li2024controlnet++} improved control precision under multiple conditions by introducing a pixel-level cycle consistency loss, enhancing realism. 
Follow Your Pose~\cite{ma2024follow} fine-tuned a diffusion model with human-centric data in the first stage and trained temporal attention layers in the second stage to achieve temporal consistency, enabling pose-controllable human video generation. Building on this, Follow Your Pose v2~\cite{xue2024follow} incorporated multiple control signals, such as depth maps and optical flow, to improve the accuracy of pose control and temporal consistency.

\subsection{Photo-Realistic Hand Gesture Generation}  
Recent approaches aiming at improving the fidelity of gesture generation can generally be grouped into two categories: single-stage methods~\cite{narasimhaswamy2024handiffuser,park2024attentionhand,fu2024adaptive} and multi-step pipelines~\cite{lu2024handrefiner, pelykh2024giving, wang2025realishuman, zhu2024mole}.

\noindent \textbf{Single-stage Methods:}  
These methods produce the entire image in one forward pass while simultaneously directing the model's focus toward learning the detailed features of the hand area. For instance, HandDiffuser~\cite{narasimhaswamy2024handiffuser} strengthens the diffusion model’s ability to capture the structural patterns of hand gestures by integrating 3D hand mesh data through a dedicated encoding module. Also leveraging the rich information from 3D mesh, a novel text-guided framework named AttentionHand~\cite{park2024attentionhand} uses 3D hand mesh prior with text and bounding box for controllable hand image generation.  
Overall, achieving realistic hand generation within a one-step framework is still a complex task due to data-intensive and the need to simultaneously maintain both holistic image coherence and intricate hand geometry.

\noindent \textbf{Multi-stage Methods:}  
These approaches divide the gesture generation task into several stages, where each stage applies a specific generative model to handle particular regions of the image. The outputs from all stages are then combined into a final result. A key step in this pipeline is localized inpainting: based on the coarse image from the initial stage, a hand-focused inpainting model is used to enhance the quality of the hand area. Compared with single-stage frameworks, multi-stage strategies can improve the realism and accuracy of hand synthesis, although they usually require more complex inference procedures and additional model components.
HandRefiner~\cite{lu2024handrefiner} adapts ControlNet~\cite{zhang2023adding} by fine-tuning it with hand depth maps, allowing the inpainting diffusion model to repair the distorted hand area. However, it still struggles to maintain consistency between the refined hand area and the surrounding content generated in the first step. 
Leveraging hand and face masks from DINO v2~\cite{oquab2024dinov2} for segmentation, RealisHuman~\cite{wang2025realishuman} introduces a diffusion model controlled by multiple prior conditions to generate realistic facial and hand areas, followed by a second step to repaint the transitions between background and modified human parts. Nevertheless, the prior used comes from a 2D projection of a single viewpoint and still does not build complete structural information.
Although previous single-stage and multi-stage approaches leverage multi-modal priors to enhance the quality of hand gesture generation, none of them incorporate complementary information from multiple viewpoints. This is particularly important for hand gestures, which often involve complex spatial structures and severe occlusions. In contrast to these methods, our proposed MUFEN extracts and fuses multi-view features along with a multi-modal feature of hand gestures. By providing a more comprehensive representation of hand gestures, our method guides the diffusion model to generate more realistic hand gestures within a single-stage inference framework, removing the need for additional stages or external inpainting processes.

\begin{figure*}[t]
  \centering
  \includegraphics[width=0.8\linewidth]{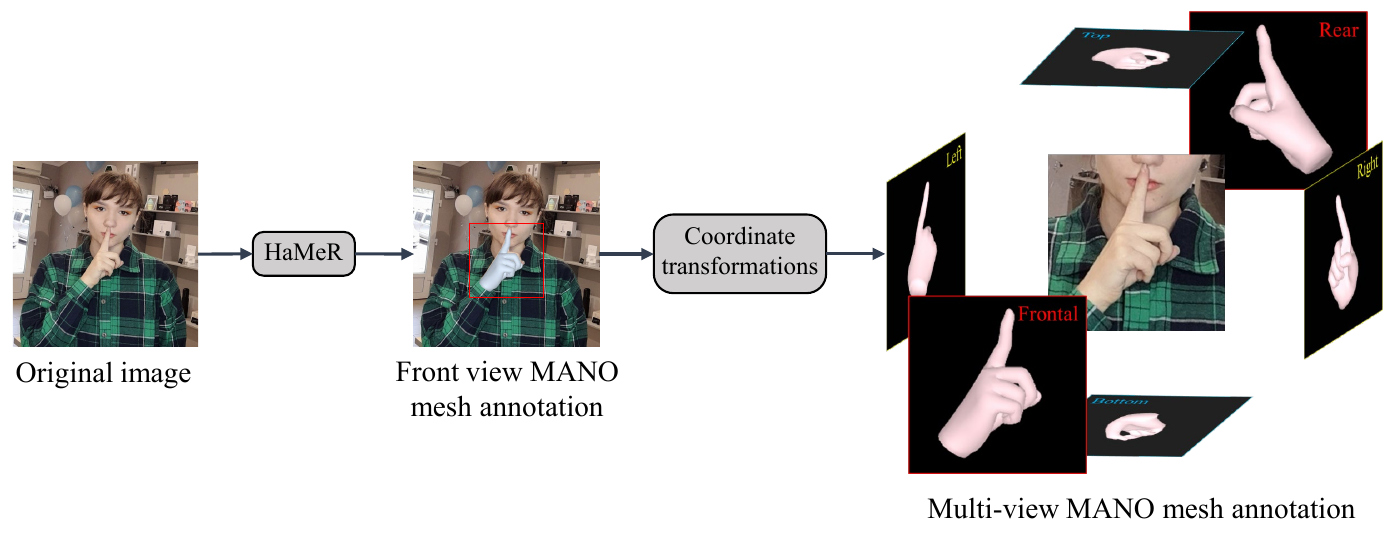}
  \caption{Multi-view hand MANO mesh rendering pipeline.}
  \label{fig:2}
\end{figure*}

\section{Multi-Modal UNet-based Feature Encoder}

\subsection{Preliminaries}

\subsubsection{Latent Diffusion Model with ControlNet Guidance}

In this paper, we utilize a latent diffusion framework as our core generative backbone due to its strong generation performance and improved efficiency. Latent Diffusion Models (LDMs)~\cite{rombach2022high} operate in a compressed latent space rather than directly in pixel space, significantly reducing computational requirements when handling high-dimensional data. During the forward diffusion stage, Gaussian noise is incrementally injected into the latent variable $\mathbf{z}_0$, producing a noisy version $\mathbf{z}_t$ according to the formula:
\begin{equation}
\mathbf{z}_t = \sqrt{\bar{\alpha}_t} \mathbf{z}_0 + \sqrt{1 - \bar{\alpha}_t} \boldsymbol{\epsilon}, \quad \boldsymbol{\epsilon} \sim \mathcal{N}(\mathbf{0}, \mathbf{I}),
\end{equation}
where $\bar{\alpha}_t$ defines the variance schedule. To reconstruct the denoised representations, the reverse process employs a neural model $\epsilon_\theta$ to estimate the original noise component and thus denoise $\mathbf{z}_t$. The model is optimized using a noise prediction loss based on mean squared error:
\begin{equation}
\mathcal{L}_{\text{LDM}} = \mathbb{E}_{\mathbf{x}, \boldsymbol{\epsilon}, t} \left[ \|\boldsymbol{\epsilon} - \epsilon_\theta(\mathbf{z}_t, t)\|_2^2 \right],
\end{equation}
where $\mathbf{z}_t$ is encoded from the input data $\mathbf{x}$ by an autoencoder. This formulation allows for efficient sampling and ensures that the decoded outputs from the latent space maintain high fidelity to the original data.

To further enhance the controllability of the generation process, especially for structure-aware image synthesis, ControlNet~\cite{zhang2023adding} is incorporated as an auxiliary module. ControlNet extends diffusion models by introducing an additional conditioning pathway that guides the generation according to specific structural cues, such as key points, depth, or semantic maps. This mechanism allows the model to follow external constraints while preserving the quality of the generated content. It does so by integrating control embeddings $c$ into the denoising network at every step of the diffusion. The objective function of ControlNet adapts the LDM loss into a conditional form:
\begin{equation}
\mathcal{L}_{\text{denoise}} = \mathbb{E}_{\mathbf{x}, \boldsymbol{\epsilon}, t, c} \left[ \|\boldsymbol{\epsilon} - \epsilon_\theta(\mathbf{z}_t, t, c)\|_2^2 \right],
\end{equation}
where the control signal $c$ provides task-specific guidance to the model.

\subsubsection{MANO Hand Representation}  

To model the 3D structure of hand gestures, we adopt the hand Model with Articulated and Non-rigid deformations (MANO) model~\cite{romero2017embodied}, which is a parametric hand model similar in design to the SMPL body model~\cite{loper2023smpl}. The MANO framework produces a three-dimensional hand mesh composed of 778 vertices, each described by a triplet of spatial coordinates \((x, y, z)\). Moreover, it provides a set of 21 semantic key points representing essential anatomical joints of the hand, including the wrist and four articulations per finger. 
For multi-view mesh rendering, we utilized the recent mesh annotator HaMeR ~\cite{pavlakos2024reconstructing} for accurate recovery of MANO mesh parameters. This procedure allowed us to extract both the MANO pose and shape parameters for each hand gesture, along with the corresponding hand bounding boxes.

\subsection{Multi-View MANO Mesh Rendering}

Figure~\ref{fig:2} illustrates our pipeline for rendering Multi-View MANO Meshes for each gesture. Specifically, for each input image, we first employ HaMeR~\cite{pavlakos2024reconstructing} to extract the meshes of all hand gestures present in the image and render front-view mesh images of these gestures. Subsequently, we adjust the mesh vertex coordinates and camera positions through coordinate transformations to achieve rendering from other viewpoints.
We begin by detailing the coordinate transformation from the front view to the rear view, which involves a 180-degree rotation around the Y-axis, effectively reversing the direction of observation. Specifically, the transformation from front view to rear view can be formalized as:
\[
\mathbf{v}_{\text{rear}} = \mathbf{R}_y(\pi) \cdot \mathbf{v}_{\text{front}},
\]
where $\mathbf{v}_{\text{front}}$ represents the vertex coordinates in the front view, $\mathbf{v}_{\text{rear}}$ represents the transformed coordinates for the rear view, and $\mathbf{R}_y(\pi)$ is the rotation matrix for a 180-degree rotation around the Y-axis:
\[
\mathbf{R}_y(\pi) = 
\begin{pmatrix}
-1 & 0 & 0 \\
0 & 1 & 0 \\
0 & 0 & -1
\end{pmatrix},
\]
This transformation effectively negates the X-coordinate: $x_{\text{rear}} = -x_{\text{front}}$, preserves the Y-coordinate: $y_{\text{rear}} = y_{\text{front}}$, and negates the Z-coordinate: $z_{\text{rear}} = -z_{\text{front}}$

$\mathbf{t}_{\text{camera\_front}} = (t_x, t_y, t_z)^T$ is the original camera translation vector in the front view. To maintain proper viewing geometry after the vertex transformation, the camera position must also be adjusted. Specifically, the X-component of the camera translation vector must be mirrored:
\[
\mathbf{t}_{\text{camera\_rear}} = 
\begin{pmatrix}
-t_x \\
t_y \\
t_z
\end{pmatrix}.
\]

This transformation preserves the relative spatial arrangement of the vertices while providing a view from the opposite side, which is essential for the comprehensive analysis of 3D hand models where both front and rear views may reveal different structural features.

To transform coordinates from the front view to the left or right side view, a rotation around the Y-axis is applied. This transformation can be represented as:
\[
\mathbf{v}_{\text{side}} = \mathbf{R}_y(\theta) \cdot \mathbf{v}_{\text{front}},
\]
where $\mathbf{v}_{\text{front}}$ and $\mathbf{v}_{\text{side}}$ represent vertex coordinates in the front and side viewpoints respectively, and $\mathbf{R}_y(\theta)$ is the rotation matrix around the Y-axis:
\[
\mathbf{R}_y(\theta) = 
\begin{pmatrix}
\cos(\theta) & 0 & \sin(\theta) \\
0 & 1 & 0 \\
-\sin(\theta) & 0 & \cos(\theta)
\end{pmatrix},
\]
where $\theta = \frac{\pi}{2}$ for the right side view and $\theta = -\frac{\pi}{2}$ for the left side view.

After this rotation, the original X-coordinate translational offset ($t_x$) becomes a Z-axis offset: $t_z = -t_x$ for right side view and $t_z = t_x$ for left side view.

To transform coordinates from the front view to the top or bottom view, a rotation around the X-axis is applied:
\[
\mathbf{v}_{\text{vertical}} = \mathbf{R}_x(\phi) \cdot \mathbf{v}_{\text{front}},
\]
where $\mathbf{v}_{\text{front}}$ and $\mathbf{v}_{\text{vertical}}$ represent vertex coordinates in the front and vertical viewpoints respectively, and $\mathbf{R}_x(\phi)$ is the rotation matrix around the X-axis:
\[
\mathbf{R}_x(\phi) = 
\begin{pmatrix}
1 & 0 & 0 \\
0 & \cos(\phi) & -\sin(\phi) \\
0 & \sin(\phi) & \cos(\phi)
\end{pmatrix},
\]
where $\phi = \frac{\pi}{2}$ for the top view and $\phi = -\frac{\pi}{2}$ for the bottom view.

After this rotation, the original Y-coordinate translational offset ($t_y$) becomes a Z-axis offset: $t_z = t_y$ for top view and $t_z = -t_y$ for bottom view.

We also adjusted the parameters to make the lighting and colors consistent across the viewpoint meshes. To ensure perceptual uniformity across all viewpoints, the rendering pipeline adopts a unified color parameterization and a standardized multi-light illumination model. The base mesh color is fixed as $\mathbf{C}_{\text{mesh}} = (1.0, 1.0, 0.9) \in \mathbb{R}^3$ with a constant background color $\mathbf{C}_{\text{bg}} = (0,0,0)$, and per-vertex colors are uniformly assigned as $(\mathbf{C}_{\text{mesh}}, 1.0)$. Left--right hand differentiation is applied via fixed chromatic offsets, ensuring consistent hue mapping across all rendered instances. Material parameters, including metallic factor $m=0.0$ and roughness $r=1.0$, remain fixed to eliminate view-dependent reflectance variations.

Illumination consistency is maintained through a fixed composite lighting model
\[
\mathbf{L}_{\text{total}} = \mathbf{L}_{\text{ambient}} + \mathbf{L}_{\text{dir}} + \mathbf{L}_{\text{point}} + \mathbf{L}_{\text{raymond}},
\]
where ambient light $(0.3,0.3,0.3)$ provides a global base tone, and directional, point, and Raymond lights\footnote {\url{https://github.com/mmatl/pyrender/blob/master/pyrender/viewer.py}} are placed at pre-defined positions with constant intensities. View-specific compensation lights are added for oblique or side views while preserving the global intensity balance, ensuring that shading gradients, color perception, and brightness remain consistent across all rendered perspectives.

These transformations collectively enable comprehensive multi-view visualization of 3D mesh models while preserving their relative spatial configurations.

\subsection{Dual Stream Rendering Encoder for the most Informative Perspective}
The information contained in the six basic view meshes rendered by the above method is complementary and grouped into three complementary pairs: (front, back), (left, right), and (top, bottom). 
The views in a complementary pair complement each other's occluded information. For different gestures, the amount of information contained in these three sets of views is also different.

Based on the characteristics of human gestures, we calculate the projected area of the six views on their view planes and select the one with the largest projected area as the most informative view for the gesture. This ensures coverage of critical gesture cues while reducing redundancy. Taking the gesture in Figure~\ref{fig:2} as an example, the most informative pair of views was calculated to be the front and rear views.

\begin{figure}[t]
  \centering
  \includegraphics[width=0.8\linewidth]{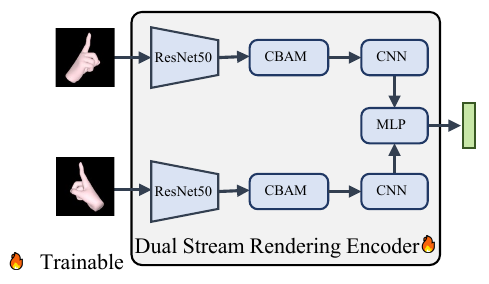}
  \caption{The dual-stream rendering encoder receives a pair of mesh images, extracts the features from them separately, and then fuses the extracted features for output.}
  \label{fig:3}
\end{figure}

Then, we design a dual-stream rendering encoder to extract and fuse this complementary information for the two rendered meshes in the most informative view. As shown in Figure~\ref{fig:3}, the dual-stream encoder integrates two rendering encoders, each processing one of the two input mesh images. The features from both encoders are concatenated and passed through the fusion network to produce a unified feature representation. 

Specifically, in each rendering encoder, a pre-trained ResNet50 \cite{he2016deep} model, truncated before the final layers, is first used to extract features. Then, a Convolutional Block Attention Module \cite{woo2018cbam} with both channel and spatial attention mechanisms is applied to enhance these features. Finally the feature channel is reduced to 1280 for output by a convolutional layer. 

Two rendering encoders are followed by a fusion network, which consists of a series of fully connected layers with ReLU activations \cite{agarap2018deep}. It also employs a residual connection to preserve the original input features while learning new representations.

\subsection{Multi-Modal UNet-based Feature Encoder}

\begin{figure}[t]
  \centering
  \includegraphics[width=\linewidth]{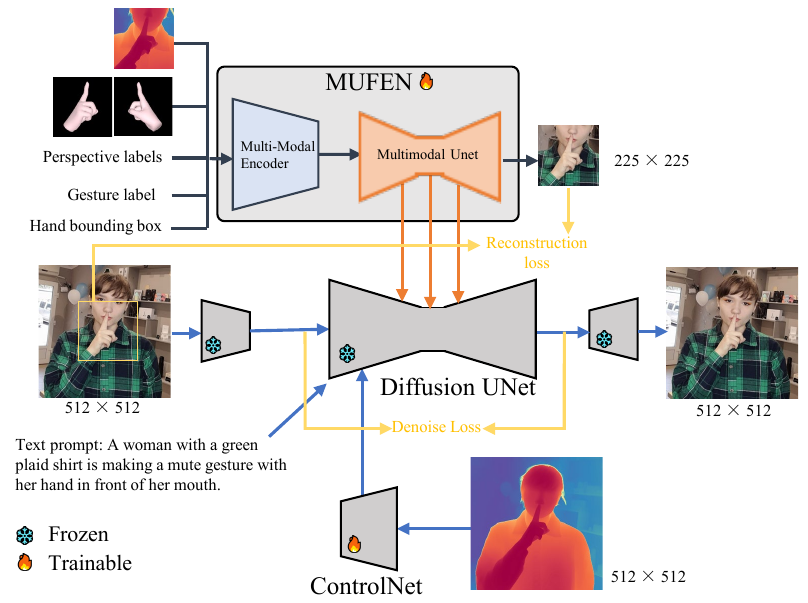}
  \caption{Training pipeline of Multi-Modal UNet-based Feature Encoder (MUFEN).}
  \label{fig:5}
\end{figure}

Figure~\ref{fig:5} demonstrates the training pipeline of Multi-Modal UNet-based Feature Encoder (MUFEN), which consists of two parts: a Multi-Modal Encoder and a Multi-Modal UNet.  This architecture is responsible for the encoding of different modalities, their fusion, and the subsequent processing through a UNet to produce a final output. 

\begin{figure}[t]
  \centering
  \includegraphics[width=\linewidth]{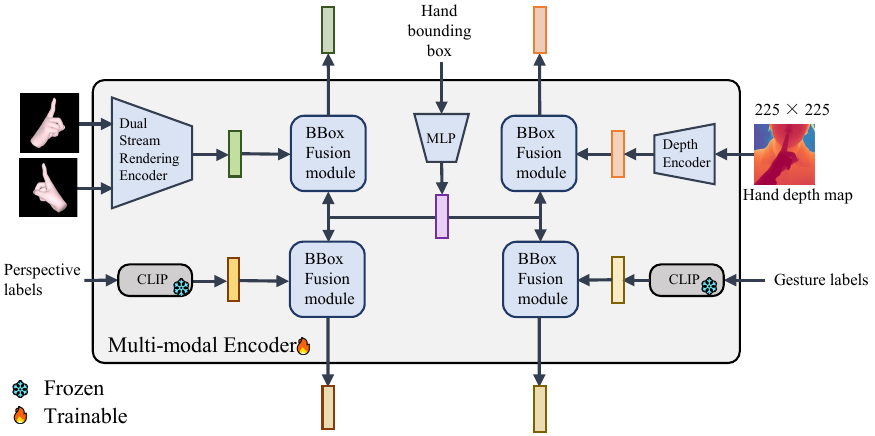}
  \caption{Our multi-modal encoder integrates features from mesh images, text, depth images, and bounding boxes with a dedicated encoder for each modality.}
  \label{fig:4}
\end{figure}

As shown in Figure~\ref{fig:4}, the Multi-Modal Encoder encodes and integrates features from different modalities, including mesh images, text, depth images, and bounding boxes. Each modality has a dedicated encoder, and bounding box features are fused into each modality via a tailored fusion mechanism. Specifically, the mesh modality is processed by a Dual Stream Rendering Encoder, which extracts features from two cropped mesh images using bounding boxes, as described in the previous section. The depth encoder shares the same structure as a single rendering encoder and extracts both gesture and environmental features from the depth map. Gesture-related text features are extracted from the text label by pre-trained openai-clip-vit-large-patch14~\cite{radford2021learning}. The Bounding Box Encoder built with multi-layer perceptrons (MLP)~\cite{kruse2022multi} and extracts gesture localization features from the bounding box coordinate data consisting of the 2D coordinates of the top-left and bottom-right corners. These localization features are later used to enhance the spatial localization ability of other modalities through BBox Fusion modules.

\begin{figure}[t]
  \centering
  \includegraphics[width=0.8\linewidth]{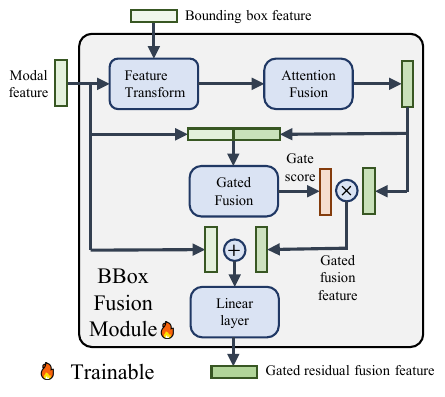}
  \caption{The bounding box (BBox) fusion module adds localization information from the gesture bounding box to the multimodal features extracted by MUFEN through attention and gating mechanisms.}
  \label{fig:6}
\end{figure}

For each modality, there is a dedicated BBox Fusion module, as shown in Figure~\ref{fig:6}. This module integrates the feature of the bounding box into the feature specific to other modalities using a combination of attention mechanisms and gating strategies. The fusion process involves: Feature Transformation, where the bounding box features are transformed to match the feature dimension of the modality; Attention Fusion, where the bounding box features act as queries, while the modality features serve as keys and values in an attention mechanism, allowing the bounding box features to capture context from the modality features; Learnable Gated Fusion, where a gating mechanism controls the extent to which the bounding box information is integrated into the modality features. By using a learnable gate, the model can adaptively determine the importance of the context features for each token and each sample in the batch, leading to more nuanced and context-aware feature representations; Residual Connection and Projection, where the final fused features are obtained by adding the original modality features to the gated context features, followed by a linear projection.

The output of the Multi-Modal Encoder is fed into a Multi-Modal UNet. Specifically, modality features fused with bounding box information are first concatenated along the feature dimension and passed through an MLP to obtain a unified fused feature, which then passes through an identity layer, preserving its shape. The spatial dimensions are reduced from 16×16 to 8×8 in the DownBlock, which includes a convolutional layer with self-attention, allowing the model to capture long-range dependencies. At the bottleneck, the features are processed with a self-attention mechanism to further enhance feature interactions without changing the spatial dimensions. During the up-sampling process, the spatial resolution is restored from 8×8 to 16×16 using bilinear interpolation, followed by a convolutional layer. Cross-attention is applied to integrate skip connections from the encoder. Finally, the features are up-scaled to the target resolution of 225×225 and generates the final RGB image output using a convolutional layer.

\subsection{Training}

As shown in Figure~\ref{fig:5}, MUFEN is trained with the following loss function:
\begin{equation} \label{E4}
\begin{aligned}
{\mathcal{L}} = \mathcal{L}_{\mathrm{denoise}} + \lambda {\mathcal{L}}_{{\rm{reHand}}},
\end{aligned}
\end{equation}
where ${{\mathcal{L}_{\mathrm{reHand}}} =} {\mathbb E} \left| {{\bf I}_{{\rm{reHand}}} - \hat {\bf I}}_{{\rm{reHand}}} \right|$, ${\bf I}_{{\rm{reHand}}}$ and ${\hat {\bf I}}_{{\rm{reHand}}}$ are the ground truth gesture region image and the reconstructed gesture region image, respectively, and $\lambda$ is a weight to balance ${\mathcal{L}_{\mathrm{denoise}}}$ and ${\mathcal{L}_{\mathrm{reHand}}}$.

\section{Experiments}
\begin{figure*}[h]
  \centering
  \includegraphics[width=0.8\linewidth]{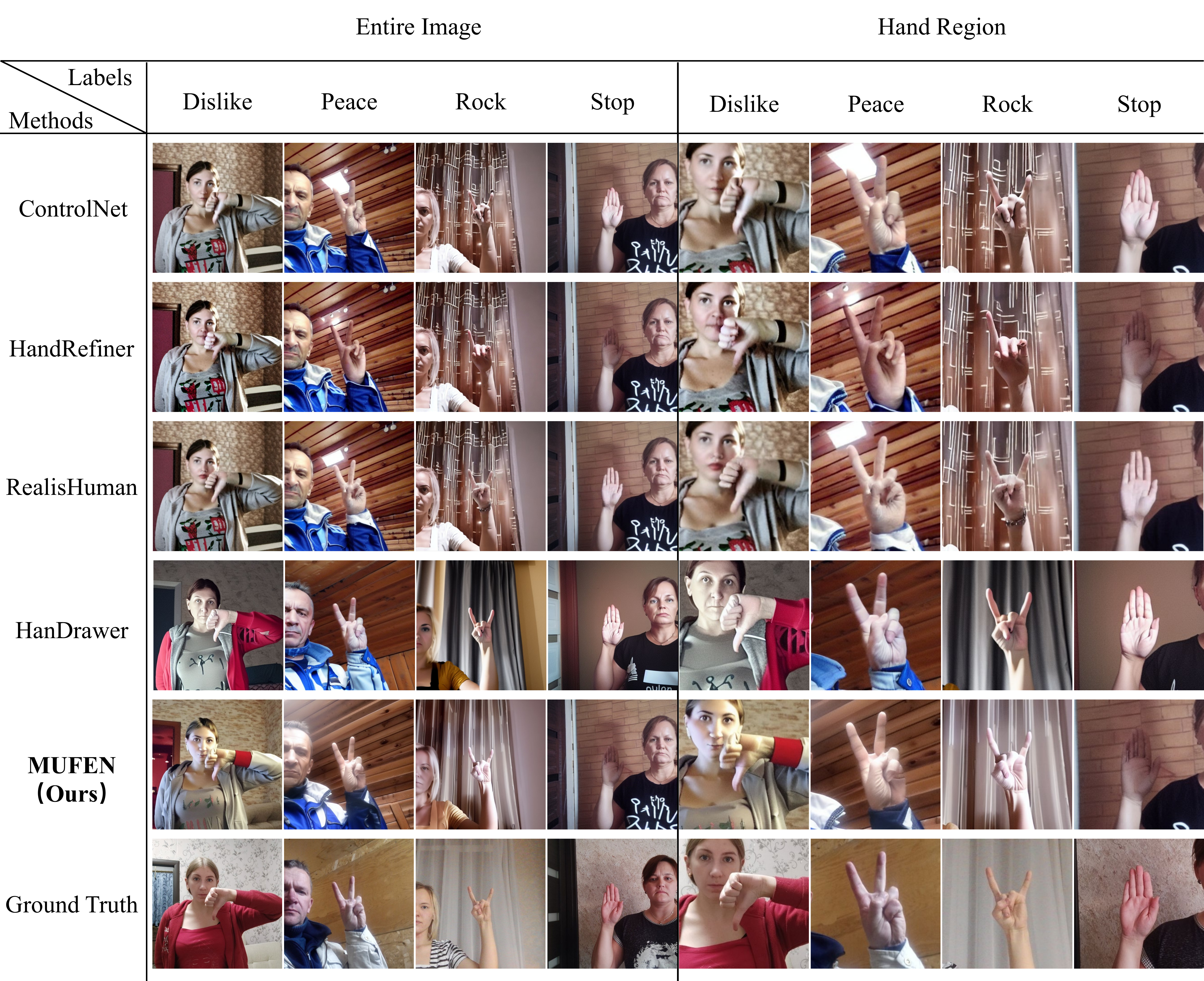}
  \caption{Qualitative results for different methods.}
  \label{fig:9}
\end{figure*}

\subsection{Implementation Details}
We used the processed HaGRID v1 dataset provided by ~\cite{fu2025handrawer}, which contained 28,814 training images and 7,623 testing images across 18 gesture categories, with the entire image size $512 \times 512$. Based on the existing multi-modal annotations, we additionally rendered rear, left, right, top, and bottom views mesh using off-the-shelf tool HaMeR~\cite{pavlakos2024reconstructing} and recorded the corresponding labels and projection areas for each view. For each gesture, we selected the two views with the largest projection areas as the complementary view pair. During training, we fixed the parameters of the Diffusion v1.5 model and trained the proposed MUFEN together with ControlNet~\cite{zhang2023adding} for 90,000 steps using a learning rate of $1 {e^{ - 6}}$ and a batch size of 6. The loss weight $\lambda$ was set to 0.1. For comparison, we used ControlNet~\cite{zhang2023adding}, HandRefiner~\cite{lu2024handrefiner}, RealisHuman~\cite{wang2025realishuman}, and HanDrawer~\cite{fu2025handrawer} as baselines. All training and inference were conducted on a single NVIDIA Tesla A100 80GB GPU.

\subsection{Evaluation Metrics}
Frechet Inception Distance (FID)~\cite{heusel2017gans} and Kernel Inception Distance (KID)~\cite{binkowski2018demystifying} are used for entire image evaluation, and hand region metrics FID-Hand and KID-Hand are used to measure the generation quality of hand regions. Hand regions are cropped from an entire image with $299\times299$ size centred on the hand bounding box, because Inception v3 model for calculating FID requires a  $299\times299$  size image as input. In all cases, lower numerical values represent better performance. FID-H and KID-H are our primary quantitative evaluation metrics as they focus on the hand regions.

\begin{table*}
  \caption{Quantitative results on HaGRID v1 dataset. MUFEN outperforms existing methods in all metrics, with substantial improvement repetition. The best performance is highlighted in \textbf{bold}.}
  \label{tab:commands}
  \begin{tabular}{l|cccc}
    \toprule
    Method  & FID-Hand ↓ & KID-Hand ↓ & FID ↓ & KID ↓ \\
    \midrule
    SD v1.5 + ControlNet~\cite{zhang2023adding} & 32.2157 & 0.0238$\pm$0.0007 & 31.4976 & 0.0238$\pm$0.0002 \\
    HandRefiner~\cite{lu2024handrefiner}  & 35.4393 & 0.0259$\pm$0.0004 & 35.7291 & 0.0296$\pm$0.0008\\
    RealisHuman~\cite{wang2025realishuman}  & 30.2902 & 0.0210$\pm$0.0004 & 31.0369 & 0.0232$\pm$0.0001\\
    HanDrawer~\cite{fu2025handrawer} & 28.7506 & 0.0201$\pm$0.0008 & 26.8279 & 0.0196$\pm$0.0002 \\
    MUFEN (Ours)  & \textbf{26.8526} & \textbf{0.0173$\pm$0.0007} & \textbf{26.7749} & \textbf{0.0194$\pm$0.0003} \\
    \bottomrule
  \end{tabular}
  \label{tab:table1}
\end{table*}

\begin{table*}
\caption{Paired \textit{t}-test results of MUFEN compared with other methods over 18 gestures. 
For each comparison, the $p$-values of both FID-Hand and KID-Hand metrics are reported, along with the number of gestures for which MUFEN achieves better performance.}
  \label{tab:commands}
  \begin{tabular}{l|cccc}
    \toprule
MUFEN compared With & FID-Hand $p$-value & Better Count & KID-Hand $p$-value & Better Count \\
\midrule
SD v1.5 + ControlNet~\cite{zhang2023adding} & $7.6\times 10^{-8}$ & 18 & $5.6\times 10^{-8}$ & 18 \\
HandRefiner~\cite{lu2024handrefiner} & $3.5\times 10^{-7}$ & 18 & $3.5\times 10^{-7}$ & 18 \\
RealisHuman~\cite{wang2025realishuman} & $0.305$ & 13 & $0.007$ & 13 \\
HanDrawer~\cite{fu2025handrawer} & $0.059$ & 12 & $0.018$ & 12 \\
\bottomrule
\end{tabular}
\label{tab:table2}
\end{table*}

\begin{table*}
  \caption{Ablation results on different modalities and number of views. MUFEN exceeds all ablated variants on hand-region metrics. The best quality is highlighted in \textbf{bold}.}
  \label{tab:commands}
  \begin{tabular}{l|cccc}
    \toprule
    Different settings  & FID-Hand ↓ & KID-Hand ↓ & FID ↓ & KID ↓ \\
    \midrule
    4 views (front, back, left, right)  & 31.92 & 0.0222 ± 0.0007 & 30.52 & 0.0230 ± 0.0005\\
    4 views (front, back, top, bottom)  & 32.03 & 0.0221 ± 0.0007 & 30.74 & 0.0235 ± 0.0004\\
    4 views (left, right, top, bottom) & 30.78 & 0.0205 ± 0.0008 & 28.79 & 0.0207 ± 0.0004 \\
    6 views  & 39.18 & 0.0297 ± 0.0009 & 37.78 & 0.0308 ± 0.0005 \\
    w/o depth map  & 38.11 & 0.0298 ± 0.0010 & 35.37 & 0.0288 ± 0.0003\\
    w/o mesh  & 31.37 & 0.0217 ± 0.0008 & 31.35 & 0.0241 ± 0.0003\\
    w/o gesture label & 27.88 & 0.0180 ± 0.0006 & 27.94 & 0.0209 ± 0.0001 \\
    w/o bbox fusion  & 27.73 & 0.0175 ± 0.0007 & \textbf{26.46} & 0.0186 ± 0.0003 \\
    MUFEN (Ours) & \textbf{26.85} & \textbf{0.0173$\pm$0.0007} & 26.77 & \textbf{0.0194$\pm$0.0003}
 \\
    \bottomrule
  \end{tabular}
  \label{tab:table3}
\end{table*}

\subsection{Quantitative Results}
The quantitative results comparing several methods are presented in Table \ref{tab:table1}. Notably, the results indicate that the proposed MUFEN method achieves the best performance across all metrics. For the hand-related metrics, MUFEN achieves an FID-Hand of 26.85 and a KID-Hand of 0.0173±0.0007, showing a substantial improvement compared to the other methods.
This comparison indicates that the proposed MUFEN method brings great improvements, specifically in the generation of hand regions. This superior performance is primarily attributed to the use of a dual-stream encoder that extracts and fuses complementary information from different viewpoints of the hand. Moreover, the carefully designed multi-modal encoder further enhances the generation capability of the hand modality. These architectural innovations collectively contribute to a more effective representation and synthesis of hand features, as clearly reflected by the improved FID-Hand and KID-Hand metrics.

Table \ref{tab:table2} illustrates the paired \textit{t}-test results of MUFEN compared with other methods over 18 gestures.  The paired t-tests across 18 gestures show MUFEN significantly outperforms ControlNet and HandRefiner ($p < 10^{-6}$, i.e. better on 18 out of 18 gestures). It also improves over HanDrawer on KID-Hand ($p = 0.018$) and is near-significant on FID-Hand ($p = 0.059$). Compared to RealisHuman, MUFEN achieves significance on KID-Hand ($p = 0.007$), indicating better detail fidelity. These results confirm MUFEN’s advantages over baselines, especially under the KID metric.

To assess computational cost, we compare inference time across single stage inference methods. On a NVIDIA Tesla A100 80GB GPU, MUFEN takes 2.39 seconds to generate an image, slightly faster than HanDrawer (2.45s), marginally slower than MUFEN-single-view (2.12s), and ControlNet is fastest (1.65s). MUFEN offers stronger generation quality with moderate cost. On a NVIDIA GeForce RTX 3080 Laptop GPU, rendering six-view meshes takes 1.45s, compared to 0.48s for front-view-only rendering.

\subsection{Qualitative Results}
Figure~\ref{fig:1} and Figure~\ref{fig:9} illustrate the results of the qualitative analysis. It is shown that the hand regions generated by our proposed MUFEN are more realistic, and the gestures are more accurate.

\subsection{Ablation Studies}
Table \ref{tab:table3} shows ablation results on different modalities and number of views. Clearly, MUFEN with original settings consistently outperforms all ablated variants on hand-region metrics, validating our design choices. Regarding the choice of number of views, using more predefined views (4 or 6) degrades performance, due to information redundancy from views overlap. While MUFEN selects the co-axial view pair with the largest combined projected area (2 views) as input, which minimizes redundancy and enhances gesture representation, achieving better performance and validating our complementary view selection strategy. Regarding the combination of multimodal, removing the depth map causes the most degradation (FID +8.60, FID-Hand +11.26), indicating its importance as a structural prior. Excluding the mesh also significantly affects performance (FID +4.58, FID-Hand +4.52), confirming its role in gesture geometry modeling. Gesture labels yield moderate gains (FID +1.17), while bounding box fusion helps localize hand regions, as reflected in increased FID-Hand (+0.88) when removed. These results affirm the importance of each modality and corresponding module in our framework and support the effectiveness of MUFEN’s multimodal design.

\section{Conclusions}
This paper proposed a diffusion model-based framework to enhance the realism and accuracy of generated hand gestures in human-centric generation tasks through multi-view priors. Specifically, we employed coordinate transformations on 3D MANO mesh rendering combined with viewpoint camera position adjustments to derive six orthogonal mesh projections (front, rear, left, right, top, bottom). We identified the optimal pair of viewpoints that contain the maximum geometric information for prior integration via projected areas of the rendered meshes. We then designed a dedicated Multi-Modal UNet-based Feature Encoder (MUFEN) to fuse the comprehensive hand modeling features from multi-view meshes and multi-modal features contained in other modalities and enhanced the accuracy of feature fusion with localization feature from the bounding box. Experiments demonstrated that the proposed framework significantly improved gesture generation quality, achieving SOTA performance across both quantitative metrics and qualitative assessments.

\section{Acknowledgments}

This work was supported by the China Scholarship Council.  It was also supported by Engineering and Physical Sciences Research Council [grant number EP/Y009800/1], through Keystone project funding from Responsible Ai UK (KP0016).  This research utilized Queen Mary's Andrena HPC facility, supported by QMUL Research-IT.  

\bibliographystyle{ACM-Reference-Format}
\bibliography{sample-base}

\end{document}